\documentclass[letter,twocolumn]{jpsj2}

\usepackage{times}
\usepackage{graphicx}% Include figure files

\title{Double-Exchange Ferromagnetism and
Orbital-Fluctuation-Induced Superconductivity \\
in Cubic Uranium Compounds}

\author{Takashi {\sc Hotta}}

\inst{Department of Physics, Tokyo Metropolitan University,
1-1 Minami-Osawa, Hachioji, Tokyo 192-0397, Japan}

\recdate{\today}

\abst{
A double-exchange mechanism for the emergence of ferromagnetism
in cubic uranium compounds is proposed
on the basis of a $j$-$j$ coupling scheme.
The idea is {\it orbital-dependent duality} of $5f$ electrons
concerning itinerant $\Gamma_8^-$ and localized $\Gamma_7^-$ states
in the cubic structure.
Since orbital degree of freedom is still active
in the ferromagnetic phase,
orbital-related quantum critical phenomenon is expected to appear.
In fact, odd-parity $p$-wave pairing compatible with
ferromagnetism is found in the vicinity of an orbital ordered phase.
Furthermore, even-parity $d$-wave pairing
with significant odd-frequency components is obtained.
A possibility to observe such exotic superconductivity in manganites
is also discussed briefly.
}

\kword{
Ferromagnetism, Orbital, Superconductivity, Uranium Compounds, Manganites
}

\begin{document}
\maketitle

%%%%% introduction %%%%%

In the Bardeen-Cooper-Schrieffer theory for superconductivity,\cite{BCS}
it was simply considered that magnetism suppresses superconductivity,
since the singlet $s$-wave electron pair mediated by phonon-induced
attraction is easily destroyed by an applied magnetic field.
However, since the pioneering discovery of superconductivity
in Ce-based heavy-fermion material \cite{Steglich}
and some uranium compounds,\cite{UBe13,URu2Si2,UPt3,UPd2Al3,UNi2Al3}
it has been gradually recognized that anisotropic superconducting
pair mediated by magnetic fluctuations generally appears
in strongly correlated electron materials.
In particular, due to successive discoveries of superconductivity
near an antiferromagnetic phase
both in $d$- and $f$-electron systems,
nowadays it is confirmed that ``magnetism is a good friend to
superconductivity''.\cite{Onuki}

When we turn our attention to the relation
between ferromagnetism and superconductivity,
it was discussed that critical magnetic fluctuations
can mediate triplet Cooper pair.\cite{Fay}
In fact, superconductivity has been observed
in a ferromagnetic phase of uranium compounds
such as UGe$_2$,\cite{UGe2} URhGe,\cite{URhGe} UIr,\cite{UIr}
and UCoGe.\cite{UCoGe}
However, for $f$-electron systems,
a microscopic theory for superconductivity has not been
satisfactorily developed so far, mainly due to
the difficulty in multi-orbital nature and strong
spin-orbit coupling of $f$ electrons.

A way to overcome such a situation is
to exploit a $j$-$j$ coupling scheme.
Along this research direction, the present author
has developed microscopic $f$-electron theories
on the basis of the $j$-$j$ coupling scheme.
\cite{Hotta1,Hotta2,Hotta3}
In the model, one $f$-electron state is characterized
by an appropriate linear combination of
the $z$ component of total angular momentum $j$.
Usually it is convenient to use the basis which diagonalizes
the crystalline electric field (CEF) potential.
In any case, we accommodate plural numbers of $f$ electrons
in such one-$f$-electron states
due to the effect of Hund's rule interaction.

For the case of cubic CEF potential, it is well known that
the $j$=5/2 sextet is split into $\Gamma_7^-$ doublet and
$\Gamma_8^-$ quartet.
Since the $\Gamma_7^-$ orbital has nodes along the cubic axes,
it has strong localized nature, while
$\Gamma_8^-$ states have itinerant nature in comparison with
$\Gamma_7^-$ electrons.
This orbital-dependent duality of $f$ electrons seems to be
a key issue of rich phenomena in $f$-electron materials.
Since electrons in localized $\Gamma_7^-$ and itinerant $\Gamma_8^-$
orbitals are coupled with the Hund's rule interaction,
we envisage a situation similar to double-exchange manganites
with mobile $e_{\rm g}$ and localized $t_{\rm 2g}$ electrons.

In this Letter, a double-exchange scenario
for the emergence of ferromagnetism in cubic uranium compounds
is proposed on the basis of the orbital-dependent duality
nature of $5f$ electrons.
We also propose some experiments to confirm
the double-exchange ferromagnetism in cubic uranium materials.
In the ferromagnetic phase, we obtain the reduced Hamiltonian
with active orbital degree of freedom.
By analyzing the model within a random phase approximation
(RPA),
we find both odd-parity $p$-wave and even-parity $d$-wave pairing
states in the vicinity of an orbital ordered state,
suggesting orbital-related quantum critical phenomena.
Finally, we briefly discuss a possibility of superconductivity in
manganites, which is well described by the double-exchange model.

%%%%% j-j coupling model %%%%%

First we briefly explain the $j$-$j$ coupling scheme.
We include the spin-orbit coupling so as to define the state
labelled by the total angular momentum $\mib{j}$,
given by $\mib{j}$=$\mib{s}$+$\mib{\ell}$,
where $\mib{s}$ and $\mib{\ell}$ are spin and angular momenta,
respectively.
For $f$-orbitals with $\ell$=3, we immediately obtain an octet with
$j$=7/2 and a sextet with $j$=5/2,
which are well separated by the spin-orbit interaction.
Since the octet level is higher than the sextet one,
it is enough to consider $j$=5/2 sextet when
local $f$-electron number is less than six.

Next we define the one $f$-electron state
in the cubic crystal structure.
It is well known that under the cubic CEF potential,
the sextet of $j$=5/2 is split into
$\Gamma_7^-$ doublet and $\Gamma_8^-$ quartet.
Note, however, that the ground state depends on the crystal structure.
For instance, in the AuCu$_3$-type cubic structure,
the energy level for $\Gamma_7^-$ doublet is lower than
that for $\Gamma_8^-$,
while for CaF$_2$-type cubic structure,
$\Gamma_8^-$ quartet becomes the ground state.
In this paper, we assume the case with $\Gamma_7^-$ ground state.

Since we consider the metallic uranium compounds,
the valance of uranium ion takes the value between three and four,
corresponding to the local $f$-electron number
between three and two.
When we accommodate two or three electrons
in $\Gamma_7^-$ and $\Gamma_8^-$ levels,
we find two possibilities of low- and high-spin states,
if we borrow the terminology of $d$-electron systems,
depending on the balance between the Hund's rule interaction
and the CEF splitting between $\Gamma_7^-$ and $\Gamma_8^-$ levels.

Readers may consider that the high-spin state
is always stabilized in $f$-electron ions,
%<<-- modify
but we should note that the effective Hund's rule interaction
$J_{\rm eff}$ in the $j$-$j$ coupling scheme
is reduced from the original Hund's rule coupling
among $f$-orbitals $J_{\rm H}$
as $J_{\rm eff}$=$J_{\rm H}(g_J-1)^2$=$J_{\rm H}/49$,\cite{Hotta1}
where $g_J$ is the Land\'e's g-factor
and $g_J$=6/7 for $J$=5/2.
%<<-- modify
In fact, we have proposed the low-spin state
for actinide ions to understand
spin and orbital structure of AnTGa$_5$ (An=U and Np;
T=Ni, Pt, Fe and Co) \cite{Hotta4,Hotta5}
and multipole order in NpO$_2$.\cite{Hotta6,Hotta7}
In this paper, on the other hand,
we attempt to find new possibility of high-spin state
concerning ferromagnetism and superconductivity.

%%%%% double-exchange scenario %%%%%

Now we discuss the $f$-electron kinetic term
in a tight-binding approximation.
When we evaluate $f$-electron hopping amplitude
$t^{\mib a}_{\tau\tau'}$ for
nearest-neighbor hopping via the $\sigma$ bond
between adjacent $f$ orbitals, it is given by
$t_{aa}^{\mib{x}}$
=$-\sqrt{3}t_{ab}^{\mib{x}}$
=$-\sqrt{3}t_{ba}^{\mib{x}}$
=$3t_{bb}^{\mib{x}}$=$3t/4$,
$t_{aa}^{\mib{y}}$
=$\sqrt{3}t_{ab}^{\mib{y}}$
=$\sqrt{3}t_{ba}^{\mib{y}}$
=$3t_{bb}^{\mib{y}}$=$3t/4$,
and $t_{bb}^{\mib{z}}$=$t$,
where indices $a$ and $b$ distinguishes two $\Gamma_8^-$ states
(see Fig.~1) and $t$ is given by $t$=$3(ff\sigma)/7$
with the use of Slater-Koster integral $(ff\sigma)$.
\cite{Slater-Koster,Takegahara}
Note that $\Gamma_7^-$ orbital is localized,
since the corresponding wavefunction has nodes
along the axis directions, as shown in Fig.~1.
On the other hand, $\Gamma_8^-$ orbitals are itinerant
and their hopping amplitudes are just the same as
those of $e_{\rm g}$ orbitals of $3d$ electrons,
\cite{Hotta2,review}
since $\Gamma_8$ is isomorphic to $\Gamma_3 \times \Gamma_6$,
where $\Gamma_3$ indicates $E$ representation
for the orbital part and $\Gamma_6$ denotes the spin part.

As mentioned above,
we assume the high-spin state in this paper.
%<<---Add
Namely, the Hund's rule interaction works
among $\Gamma_7^-$ and $\Gamma_8^-$ orbitals.
Note that Coulomb interaction in $\Gamma_7^-$ states
is larger than those for $\Gamma_8^-$ ones
in the order of $J_{\rm eff}$.
The difference of the magnitude of Coulomb interaction
between itinerant and localized orbitals is not significant
in comparison with $d$-electron systems, but
%<<---Add
in the combination with the orbital dependent duality nature,
we arrive at the {\it double-exchange model},
which is used as a canonical model for manganites.
\cite{Hotta2,review}
In this model, in order to gain the kinetic energy,
the ferromagnetic phase appears,
which is called the double-exchange ferromagnetism.
This is established in the qualitative understanding of
ferromagnetism in manganites.

%%%%%%%%%%%%%%%%%%%%%%% Fig. 1 %%%%%%%%%%%%%%%%%%%%%%%%%%%
\begin{figure}[t]
\label{fig1}
\centering
\includegraphics[width=7.5truecm]{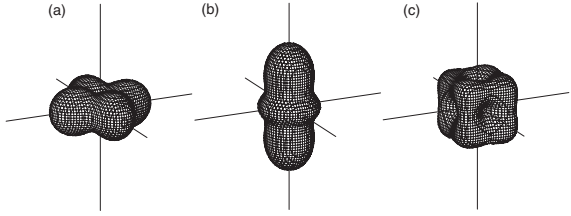}
\caption{
Charge distribution of
(a) $\Gamma_8^{-a}$, (b) $\Gamma_8^{-b}$,
and (c) $\Gamma_7$ states.
}
\end{figure}
%%%%%%%%%%%%%%%%%%%%%%%%%%%%%%%%%%%%%%%%%%%%%%%%%%%%%%%%%%

%%%%% comments on possible experiments %%%%%

In order to confirm the emergence of double-exchange ferromagnetism
in uranium compounds, we propose a couple of experiments
in analogy with manganites.
One is the observation of large negative magnetoresistance phenomenon.
Since electrons can move smoothly in the ferromagnetic phase
in comparison with the paramagnetic one,
the resistivity is drastically decreased,
when we apply a magnetic field on double-exchange materials.
The large negative magnetoresistance in cubic uranium compounds
may be an evidence for the double-exchange mechanism.
For instance, in $\beta$-US$_2$, large magnetoresistance phenomenon
has been observed,\cite{Ikeda}
although this material does not have cubic structure.

Another is more direct evidence for the relation between the Curie
temperature $T_{\rm C}$ and the kinetic energy.
In manganites with relatively wide bandwidth for conduction electrons,
it has been observed that $T_{\rm C}$ is increased
with the hole doping.\cite{review}
Since the double-exchange ferromagnetism occurs so as to
gain the kinetic energy, the ferromagnetic transition occurs
more easily when electrons can move smoothly.
Thus, we propose the appearance of the ferromagnetic metallic phase
due to the application of hydrostatic pressure or
the hole doping on insulating and/or antiferromagnetic states
of cubic uranium compounds.
In the case of uranium compounds, hole doping can be done
by thorium substitution.
It is a drastic phenomenon that ferromagnetism appears
due to thorium doping into antiferromagnetic uranium compounds.

When $\Gamma_8^-$ is lower than $\Gamma_7^-$
with large Hund's rule coupling,
the double-exchange ferromagnetism occurs
for itinerant $\Gamma_7^-$ and localized $\Gamma_8^-$.
Such a situation is realized in Nd-based filled skutterudite compounds,
in which ferromagnetism is frequently observed.
It may be interesting to seek for evidence of
double-exchange ferromagnetism in such materials.
%<<--- Add
Note also that the present mechanism cannot be directly applied to
the tetragonal Uranium material, but it works even
in the tetragonal system, when the CEF level splitting
among Kramers doublets are less than $J_{\rm eff}$ and
the lower level has localized nature.
%<<--- Add

%%%%% Reduced Hamiltonian %%%%%

Let us discuss the superconductivity in the ferromagnetic phase.
For the purpose, we consider the spinless $f$-electron model
with active orbital degree of freedom as
\begin{equation}
  H=\sum_{\mib{i,a},\tau,\tau'}t^{\mib a}_{\tau\tau'}
  f^{\dag}_{{\mib i}\tau}f_{\mib{i}+\mib{a}\tau'}
  +U\sum_{\mib{i}}n_{\mib{i}a}n_{\mib{i}b},
\end{equation}
where $f_{\mib{i}\tau}$ is the annihilation operator
for an $f$-electron in the $\tau$-orbital
of $\Gamma_8^-$ at site ${\mib i}$,
$n_{\mib{i}\tau}$=$f^{\dag}_{\mib{i}\tau}f_{\mib{i}\tau}$,
and $U$ is the inter-orbital Coulomb interaction.
Throughout this paper, we set $U$=$4t$, which is less than
the bandwidth $6t$.

Note that if $t_{\tau \tau'}^{\mib{a}}$=$t\delta_{\tau \tau'}$,
$H$ is equivalent to the well-known Hubbard model
and we simply deduce that $d$-wave superconductivity appears
near the antiferro orbital-ordered phase.
However, in actuality,
electrons hop among different adjacent orbitals.
The type of superconductivity in such a realistic
multiorbital system has been discussed actively,
\cite{Hotta8,Hotta9,Hotta10}
and quite recently, it has attracted much attention
due to the discovery of Fe-based superconductors.\cite{Ishida}

%%%%% normal Green's function %%%%%

The non-interacting Green's function ${\hat G}$ is given by
\begin{equation}
  {\hat G}^{-1}(k)=
  \left(
   \begin{array}{cc}
     i\omega_n+\mu-\varepsilon_{\mib{k}aa} & -\varepsilon_{\mib{k}ab} \\
     -\varepsilon_{\mib{k}ba} & i\omega_n+\mu-\varepsilon_{\mib{k}bb} \\
   \end{array}
  \right),
\end{equation}
where we introduce the abbreviation
$k$=$(\mib{k},i \omega_n)$,
$\mib{k}$ is the momentum,
$\omega_n$=$(2n+1)\pi T$ is the fermion Matsubara frequency
with an integer $n$ and a temperature $T$,
$\varepsilon_{\mib{k}aa}$
=$3t(\cos k_x+\cos k_y)/2$,
$\varepsilon_{\mib{k}bb}$
=$t(\cos k_x+\cos k_y+4\cos k_z)/2$,
$\varepsilon_{\mib{k}ab}$
=$\varepsilon_{\mib{k}ba}$
=$-\sqrt{3}t(\cos k_x-\cos k_y)/2$,
and a chemical potential $\mu$ controls
the $\Gamma_8^-$ electron number $\langle n \rangle$.
Since $\Gamma_7^-$ electron is assumed to be localized,
the cases of $\langle n \rangle$=1 and 2 correspond to
U$^{4+}$ and U$^{3+}$ ions, respectively.

%%%%% Gap equation %%%%%

In order to discuss superconductivity,
we solve the linearized gap equation
for anomalous self-energy ${\hat \phi}$, given by
\begin{equation}
  \phi_{\tau_1 \tau_2}(k)=-T\sum_{n'}\sum_{\mib{k}',\tau'_1,\tau'_2}
  K_{\tau_1 \tau_2, \tau'_1 \tau'_2}(k,k')
  \phi_{\tau'_1 \tau'_2}(k'),
\end{equation}
where ${\hat K}(k,k')={\hat V}(k,k'){\hat G}(k'){\hat G}(-k')$
and ${\hat V}$ is given by
\begin{equation}
 \begin{split}
   {\hat V}(k,k') & = {\hat J}
   +{\hat J} {\hat \chi}(k-k')
    [{\hat I}-{\hat J}{\hat \chi}(k-k')]^{-1}{\hat J} \\
   &+ {\hat L}-{\hat L} {\hat \chi}(k+k')
    [{\hat I}+{\hat L}{\hat \chi}(k+k')]^{-1}{\hat L}.
 \end{split}
\end{equation}
Here $J_{ab,ab}$=$J_{ba,ba}$=$L_{aa,bb}$=$L_{bb,aa}$=$U$,
${\hat I}$ denotes unit matrix,
and $\chi_{\tau_1 \tau_2, \tau_3 \tau_4}(q)$
=$-\sum_k G_{\tau_1 \tau_3}(k+q) G_{\tau_4 \tau_2}(k)$.
Here $q$=$(\mib{q},\nu_n)$, $\mib{q}$ is the momentum,
and $\nu_n$=$2n \pi T$ is the boson Matsubara frequency.
In the calculation, we use a 32$\times$32$\times$32 lattice
and 1024 Matsubara frequencies.

%$K_{\tau_1 \tau_2, \tau'_1 \tau'_2}(k,k')$
%=$\sum_{\tau_3,\tau_4}V_{\tau_1 \tau_3, \tau_4 \tau_2}(k,k')
%G_{\tau_3 \tau'_1}(k')G_{\tau_4 \tau'_2}(-k')$

%%%%% Results %%%%%

In Fig.~2, we show the phase diagram in the $(\mu, T)$ plane.
The boundary curve is determined from
the divergence in the RPA susceptibility.
The inset shows the whole phase diagram:
In the region I (0$<\mu/t<$0.85),
the orbital ordered state appears.
The ordering vector is $\mib{Q}$=$(\pi, \pi, \pi)$ at $\mu$=0,
but it is changed as $(\pi, \pi, \delta)$,
where $\delta$ is monotonically decreased
with the decrease of $\mu$
and it eventually becomes zero for $\mu/t>$0.7.
In the narrow region II (0.85$<\mu/t<$1.05),
we find $\mib{Q}$=$(\delta, \delta, \pi)$
with $\delta$=$11\pi/16$.
In the region III (1.05$<\mu/t<$1.76),
$\mib{Q}$=$(\delta, \delta, \delta)$,
where $\delta$=$\pi$ for $T/t>$0.1,
while $\delta<\pi$ for $T/t<$0.1.

%%%%%%%%%%%%%%%%%%%%%%% Fig. 2 %%%%%%%%%%%%%%%%%%%%%%%%%%%
\begin{figure}[t]
\label{fig2}
\centering
\includegraphics[width=7.5truecm]{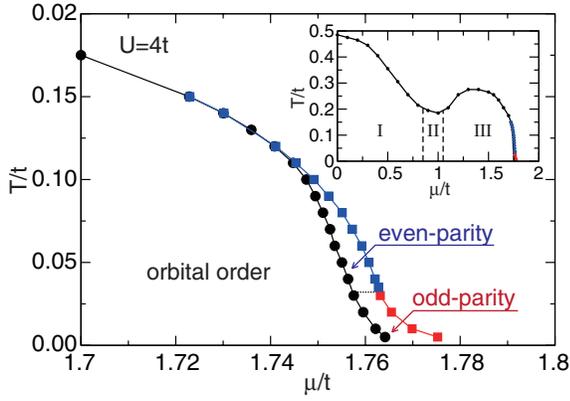}
\caption{(Color online)
Phase diagram for $U$=$4t$ near the quantum critical region.
Inset shows the whole phase diagram.
}
\end{figure}
%%%%%%%%%%%%%%%%%%%%%%%%%%%%%%%%%%%%%%%%%%%%%%%%%%%%%%%%%%

%%%%%%%%%%%%%%%%%%%%%%% Fig. 3 %%%%%%%%%%%%%%%%%%%%%%%%%%%
\begin{figure}[t]
\label{fig3}
\centering
\includegraphics[width=7.5truecm]{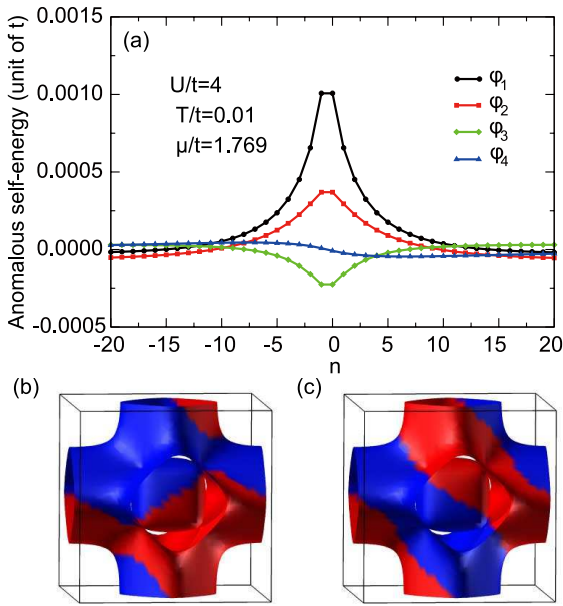}
\caption{(Color online)
(a) Anomalous self-energies with odd-parity vs. $n$
at $\mib{k}$=($\pi/3$, $\pi/2$, $3\pi/5$).
(b) Sign of $\phi_{1}(\mib{k},i\pi T)$ with odd-parity
on the Fermi surfaces.
Red and blue colors denote plus and minus signs, respectively.
(c) Sign of $\phi_{3}(\mib{k},i\pi T)$ with odd-parity
on the Fermi surfaces.
}
\end{figure}
%%%%%%%%%%%%%%%%%%%%%%%%%%%%%%%%%%%%%%%%%%%%%%%%%%%%%%%%%%

On the analogy of anisotropic superconductivity
near an antiferromagnetic critical point,
we expect the appearance of superconductivity
when the orbital order is suppressed.
In the present case, as shown in Fig.~2,
there appears superconducting pairing state
due to orbital fluctuations with $(\delta, \delta, \delta)$
around at a quantum critical point $\mu/t \approx 1.76$.
Note that orbital is $not$ the conserved quantity,
since there exists non-zero hopping amplitude
between different orbitals.
Thus, it is meaningless to define orbital singlet
and triplet by analogy with spin singlet and triplet
in the standard single-orbital Hubbard model.
Here the superconducting pair is classified only by parity.
In fact, we find that the superconducting state is
labelled by even- and odd-parity, as shown in Fig.~2.

We remark that even- and odd-frequency components are
mixed in the present case.
In order to understand this point,
it is convenient to redefine the anomalous self-energy as
$\phi_{1}(k)$=$\phi_{aa}(k)$,
$\phi_{2}(k)$=$\phi_{bb}(k)$,
$\phi_{3}(k)$=$[\phi_{ab}(k)+\phi_{ba}(k)]/\sqrt{2}$,
and
$\phi_{4}(k)$=$[\phi_{ab}(k)-\phi_{ba}(k)]/\sqrt{2}$.
First we note that the relation
$\phi_{j}(k)$=$-\phi_{j}(-k)$
always holds for $j$=1$\sim$4,
since it is due to the fermion property.
The odd-parity solutions are characterized by
$\phi^{\rm o}_{i}(\mib{k},i\omega_n)$
=$-\phi^{\rm o}_{i}(-\mib{k},i\omega_n)$
=$\phi^{\rm o}_{i}(\mib{k},-i\omega_n)$
for $i$=1$\sim$3
and
$\phi^{\rm o}_{4}(\mib{k},i\omega_n)$
=$-\phi^{\rm o}_{4}(-\mib{k},i\omega_n)$
=$-\phi^{\rm o}_{4}(\mib{k},-i\omega_n)$.
Note that $\phi_4^{\rm o}(k)$ has odd-frequency property.
On the other hand,
the even-parity solutions are characterized by
$\phi^{\rm e}_{i}(\mib{k},i\omega_n)$=
$\phi^{\rm e}_{i}(-\mib{k},i\omega_n)$=
$-\phi^{\rm e}_{i}(\mib{k},-i\omega_n)$
for $i$=1$\sim$3 and 
$\phi^{\rm e}_{4}(\mib{k},i\omega_n)$=
$\phi^{\rm e}_{4}(-\mib{k},i\omega_n)$=
$\phi^{\rm e}_{4}(\mib{k},-i\omega_n)$.
Note that $\phi_i^{\rm e}(k)$ with $i$=1$\sim$3
have odd-frequency properties.

%%%%%%%%%%%%%%%%%%%%%%% Fig. 4 %%%%%%%%%%%%%%%%%%%%%%%%%%%
\begin{figure}[t]
\label{fig4}
\centering
\includegraphics[width=7.5truecm]{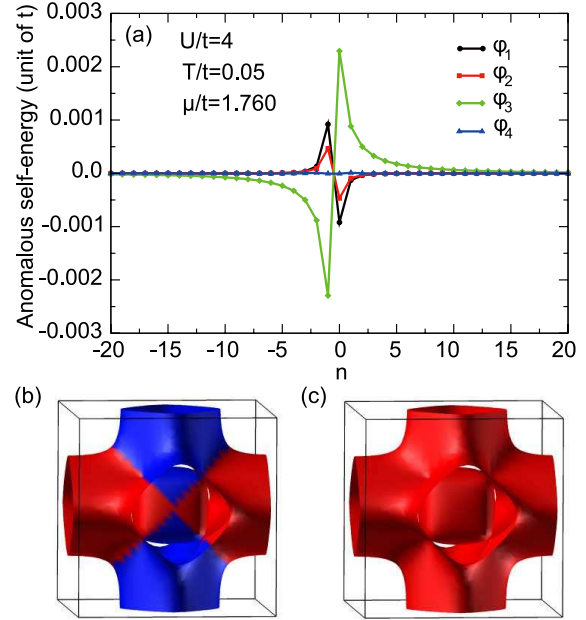}
\caption{(Color online)
(a) Anomalous self-energies with even-parity vs. $n$
at $\mib{k}$=($\pi/3$, $\pi/2$, $3\pi/5$).
(b) Sign of $\phi_{1}(\mib{k},i\pi T)$ with even-parity
on the Fermi surfaces.
Red and blue colors denote plus and minus signs, respectively.
(c) Sign of $\phi_{3}(\mib{k},i\pi T)$ with even-parity
on the Fermi surfaces.
}
\end{figure}
%%%%%%%%%%%%%%%%%%%%%%%%%%%%%%%%%%%%%%%%%%%%%%%%%%%%%%%%%%

Let us first examine the odd-parity solution
in the low-temperature region.
In Fig.~3(a), we plot $\phi_{i}$'s vs. $n$ of $\omega_n$.
As mentioned above,
$\phi_{i}$'s for $i$=1$\sim$3 are even-frequency
functions, while $\phi_{4}$ is odd-frequency one.
The absolute value of $\phi_{4}$ is relatively smaller
than those of the even-frequency ones.
Since $\phi_4(k)$ is exactly equal to the amplitude for
the antisymmetric pair of electrons on different Fermi surfaces,
the contribution of $\phi_4$ is suppressed.
In Figs.~3(b) and 3(c), we plot the signs of
$\phi_{1}(\mib{k},i\pi T)$ and $\phi_{3}(\mib{k},i\pi T)$,
respectively, on a couple of Fermi surfaces
in the first Brillouin zone.
From the results, the pairing symmetry is found to be $p$-wave.
We do not show the results for $\phi_{2}$ and $\phi_{4}$,
but $\phi_{2}$ is similar to $\phi_{1}$ and
the magnitude of $\phi_{4}$ is small
in comparison with other components.
The node positions of $\phi_3$ are different from those of
$\phi_1$, but it is due to the difference in local
symmetry of $\Gamma_8^{-a}$ and $\Gamma_8^{-b}$.
In fact, we find
$\phi_3$$\sim$$\phi_1 (\cos k_x - \cos k_y)$.

In Fig.~4, we show the results for even-parity solution
in the high-temperature region.
The $n$ dependence is depicted in Fig.~4(a).
We find that $\phi_{i}$'s for $i$=1$\sim$3
are odd-frequency functions, while
$\phi_{4}$ is even-frequency one.
Also in this case,
the contribution of $\phi_4$ is relatively small
in comparison with other components.
Thus, the even-parity solution is characterized by
the odd-frequency components,
leading to a way to observe peculiar odd-frequency
pairing.\cite{Tanaka}
From Fig.~4(b), the gap function is found to be characterized by
$d$-wave.
As observed in Fig.~4(c), $\phi_3$ seems to be $s$-wave,
due to the relation of
$\phi_3$$\sim$$\phi_1 (\cos k_x - \cos k_y)$.

Note that $\mu$ is related to the valence of uranium ion.
The critical point of $\mu/t$$\approx$1.76
corresponds to $\langle n \rangle$$\approx$1.6,
i.e., U$^{3.4+}$.
The width of the superconducting region
for the valence of uranium ion is the order of 0.01.
Namely, the region is limited, but the value
in the middle of U$^{3+}$ and U$^{4+}$ is realistic
for actual uranium metallic compounds.
Thus, we belive that the superconductivity induced
by orbital fluctuations could appear
in ferromagnetic cubic uranium compounds.

%%%%% manganites %%%%%

Here we mention a possibility to
apply the theory to manganites,
which are well described by the double-exchange model.
The superconducting region corresponds to
$\langle n \rangle$$\approx$1.6,
which denotes $e_{\rm g}$ electron number for manganites.
The situation indicates 0.4 electrons per manganese ion
form the particle-hole symmetry.
Thus, the situation is close to the half-doped manganites
with orbital ordering.
In cubic manganites with relatively wide bandwidth,
the metallic ferromagnetic phase
is known to appear near the orbital ordering.
The pattern of orbital ordering is different from the present one,
but it is expected to observe superconductivity in manganites
with high quality near half-doping.
We note that the superconducting transition temperature $T_{\rm c}$
in Fig.~2 seems to be higher than that of
the single-band Hubbard model within the same RPA.
The stabilization of the even-parity solution due to significant
odd-frequency components seems to be relevant to
the increase $T_{\rm c}$.
Since this point may open a new route to high-$T_{\rm c}$
materials, further investigations will be required in future.

%%%%% comment %%%%%

Five comments are in order.
(1) We have discussed orbital ordering and
superconductivity in the ferromagnetic phase,
but in order to confirm that the Curie temperature
is higher than the orbital-ordering temperature and
$T_{\rm c}$,
it is necessary to estimate the magnitude of Coulomb
interaction among $f$ orbitals.
This point is out of the scope of this paper,
but it is one of future problems.
(2) We have pointed out that $\Gamma_7^-$ becomes localized orbital
when we take into account only $\sigma$ bond for $f$ electron hopping.
In general, hopping amplitudes through $\pi$ and $\phi$ bonds appear
and effective hoppings through ligand anions exist.
Thus, $\Gamma_7^-$ is not perfectly localized in actual systems.
However, we still believe that orbital dependent duality
has an important starting point for the discussion 
on ferromagnetism and superconductivity.
(3) We have ignored normal self-energy effects,
but it is possible to include them, for instance,
in the fluctuation-exchange (FLEX) approximation.
Without considering the vertex corrections,
it overestimates the normal self-energy effect
such as damping of quasi-particle,
but in future, we can perform the FLEX calculation
in the combination with dynamical mean-field approximation.
(4) We have discussed superconductivity in the ferromagnetic phase
from a microscopic viewpoint, but in actuality,
it is necessary to consider how magnetic flux penetrates the system.
If the magnetic flux forms some pattern such as the Abrikosov lattice,
it indicates the ordering of localized $\Gamma_7^-$ electrons
carrying magnetic moments.
This point may lead to an interesting possibility of the coupling
between flux-lattice formation and spin-orbital order.
%<<--- Add
(5) We have proposed the spinless model,
but from a realistic viwpoint, we should include both majority and
minority spin bands.
However, the minority spin band is virtually ignored,
when minority spin density is so small that the intra-orbital
Coulomb repulsion is effectively reduced in comparison with
inter-orital Coulomb inetraction, indicating that
orbital fluctuations dominate spin ones.
%<<-- Add

%%%%% Summary %%%%%

In summary, we have proposed the double-exchange scenario
for the emergence of ferromagnetism in cubic uranium compounds.
We have found orbital-related quantum critical phenomena
such as odd-parity $p$-wave and even-parity $d$-wave
superconducting states in the vicinity of orbital-ordered phase.
This orbital-fluctuation-induced superconductivity
is expected to be found in ferromagnetic cubic uranium compounds
and cubic perovskite manganites near the half-doping.

%%%%% Acknowledgement %%%%%

The author thanks Y. Aoki, Y. Haga, R. Higashinaka, S. Kambe,
T. Maehira, and H. Sato for discussions.
This work has been supported by a Grant-in-Aid
for Scientific Research on Innovative Areas ``Heavy Electrons''
(No. 20102008) of The Ministry of Education, Culture, Sports,
Science, and Technology, Japan.
The computation in this work has been done
using the facilities of the Supercomputer Center of
Institute for Solid State Physics, University of Tokyo.

%%%%%%%%%%%%%%%%%%%%%%%%%%%%%%%%%%%%%%%%%%%%%%%%%%%%%%%%%%%%%%%%%%%%%%%
%	references
%%%%%%%%%%%%%%%%%%%%%%%%%%%%%%%%%%%%%%%%%%%%%%%%%%%%%%%%%%%%%%%%%%%%%%%


\begin{thebibliography}{99}

\bibitem{BCS}
J. Bardeen, L. N. Cooper and J. R. Schrieffer:
Phys. Rev. {\bf 108} (1957) 1175.

\bibitem{Steglich}
F. Steglich {\it et al.}:
%J. Aarts, C. D. Bredl, W. Lieke, D. Meschede,
%W. Franz and H. Sch\"afer:
Phys. Rev. Lett. {\bf 43} (1979) 1892.

\bibitem{UBe13}
H. R. Ott {\it et al.}:
%H. Rudigier, Z. Fisk and J. L. Smith:
Phys. Rev. Lett. {\bf 50} (1983) 1595.

\bibitem{URu2Si2}
T. T. M. Pastra {\it et al.}:
%A. A. Menovsky, J. van den Berg, A. J. Dirkmaat,
%P. H. Kes, G. J. Nieuwenhuys and J. A. Mydosh:
Phys. Rev. Lett. {\bf 55} (1985) 2727.

\bibitem{UPt3}
G. R. Stewart {\it et al.}:
%Z. Fisk, J. O. Willis and J. L. Smith:
Phys. Rev. Lett. {\bf 52} (1984) 679.

\bibitem{UPd2Al3}
C. Geibel {\it et al.}:
%S. Thies, D. Kaczorowski, A. Mehner, A. Grauel, B. Seidel,
%U. Ahlheim, R. Helfrich, K. Petersen, C. D. Bredl and F. Steglich:
Z. Phys. B {\bf 83} (1991) 305 .

\bibitem{UNi2Al3}
C. Geibel {\it et al.}:
%C. Schank, S. Thies, H. Kitazawa, C. D. Bredl, A. Bohm, M. Rau,
%A. Grauel, R. Caspary, R. Helfrich, U. Ahlheim, G. Weber and  F. Steglich:
Z. Phys. B {\bf 84} (1991) 1.

\bibitem{Onuki}
Y. \=Onuki: JPSJ Online -- News and Comments [November 10, 2008].

\bibitem{Fay}
D. Fay and J. Appel: Phys. Rev. B {\bf 22} (1980) 3173.

\bibitem{UGe2}
S. S. Saxena {\it et al.}:
%P. Agarwal, K. Ahilan, F. M. Grosche, R. K. W. Haselwimmer,
%M. J. Steiner, E. Pugh, I. R. Walker, S. R. Jullian, P. Monthoux,
%G. G. Lonzarich, A. Huxley, I. Sheikin, D. Braithwaite and J. Flouquet:
Nature (London) {\bf 406} (2000) 587.

\bibitem{URhGe}
D. Aoki {\it et al.}:
%A. Huxley, E. Ressouche, D. Braithwaite, J. Flouquet,
%J. -P. Brison, E. Lhotel and C. Paulsen:
Nature (London) {\bf 413} (2001) 613.

\bibitem{UIr}
T. Akazawa {\it et al.}:
%H.  Hidaka, H. Kotegawa, T. C. Kobayashi, T. Fujiwara,
%E. Yamamoto, Y. Haga, R. Settai and Y. \=Onuki:
J. Phys. Soc. Jpn. {\bf 73} (2004) 3129.

\bibitem{UCoGe}
N. T. Huy {\it et al.}:
%A. Gasparini, D. E. de Nijs, Y. Huang, J. C. P. Klaasse,
%T. Gortenmulder, A. de Visser, A. Hamann, T. G\"orlach and H. v. L\"ohneysen:
Phys. Rev. Lett. {\bf 99} (2007) 067006.

\bibitem{Hotta1}
T. Hotta and K. Ueda: Phys. Rev. B {\bf 67} (2003) 104518.

\bibitem{Hotta2}
T. Hotta: Rep. Prog. Phys. {\bf 69} (2006) 2061.

\bibitem{Hotta3}
T. Hotta and H. Harima: J. Phys. Soc. Jpn. {\bf 75} (2006) 124711.

\bibitem{Hotta4}
T. Hotta: Phys. Rev. B {\bf 70} (2004) 054405.

\bibitem{Hotta5}
H. Onishi and T. Hotta: New J. of Phys. {\bf 6} (2004) 193.

\bibitem{Hotta6}
K. Kubo and T. Hotta: Phys. Rev. B {\bf 71} (2005) 140404(R).

\bibitem{Hotta7}
K. Kubo and T. Hotta: Phys. Rev. B {\bf 72} (2005) 144401.

\bibitem{Slater-Koster}
J. C. Slater and G. F. Koster: Phys. Rev. {\bf 94} (1954) 1498.

\bibitem{Takegahara}
K. Takegahara, Y. Aoki and A. Yanase:
J. Phys. C, Solid St. Phys. {\bf 13} (1980) 583.

\bibitem{review}
E. Dagotto, T. Hotta and A. Moreo:
Phys. Rep. {\bf 344} (2001) 1.

\bibitem{Ikeda}
S. Ikeda {\it et al.}:
%H. Sakai, N. Tateiwa, T. D. Matsuda, D. Aoki, Y. Homma,
%E. Yamamoto, A. Nakamura, Y. Shiokawa, Y. Ota, K. Sugiyama, M. Hagiwara,
%K. Kindo, K. Matsubayashi, M. Hedo, Y. Uwatoko, Y. Haga and Y. \=Onuki:
%to appear in J. Phys. Soc. Jpn.
J. Phys. Soc. Jpn. {\bf 78} (2009) 114704.

\bibitem{Hotta8}
T. Takimoto {\it et al.}:
%T. Hotta, T. Maehira and K. Ueda:
J. Phys.: Condens. Matter {\bf 14} (2002) L369.

\bibitem{Hotta9}
T. Takimoto, T. Hotta and K. Ueda:
Phys. Rev. B {\bf 69} (2004) 104504.

\bibitem{Hotta10}
K. Kubo and T. Hotta:
J. Phys. Soc. Jpn. {\bf 75} (2006) 083702.

\bibitem{Ishida}
K. Ishida, Y. Nakai and H. Hosono:
J. Phys. Soc. Jpn. {\bf 78} (2009) 062001.

\bibitem{Tanaka}
K. Shigeta {\it et al.}:
%S. Onari, K. Yada and Y. Tanaka:
Phys. Rev. B {\bf 79} (2009) 174507 and references therein.

\end{thebibliography}
\end{document}